\documentclass[pdflatex,sn-nature]{sn-jnl}

\usepackage{graphicx}%
\usepackage{multirow}%
\usepackage{amsmath,amssymb,amsfonts}%
\usepackage{amsthm}%
\usepackage{mathrsfs}%
\usepackage[title]{appendix}%
\usepackage{xcolor}%
\usepackage{textcomp}%
\usepackage{manyfoot}%
\usepackage{booktabs}%
\usepackage{algorithm}%
\usepackage{algorithmicx}%
\usepackage{algpseudocode}%
\usepackage{listings}%

\usepackage{textgreek}

\theoremstyle{thmstyleone}%
\theoremstyle{thmstyletwo}%

\theoremstyle{thmstylethree}%
\newcommand{\MoO}{\textalpha-MoO\textsubscript{3}}
\newcommand{\VO}{\textalpha-\!V\textsubscript{2}O\textsubscript{5}}

\newcommand{\um}{\,\textmu m} 
\newcommand{\degree}{$^{\circ}$}

\makeatletter
\newcommand*{\rom}[1]{\expandafter\@slowromancap\romannumeral #1@}
\makeatother

\begin{document}

\makeatletter
\@twosidefalse
\makeatother

\title[Mid-IR chirality and chiral thermal emission through twisting]{Mid-IR
  chirality and chiral thermal emission through twisting}

\author*[1]{\fnm{Michael T.} \sur{Enders} \email{michael.enders@icfo.eu}}
\author[1]{\fnm{Mitradeep} \sur{Sarkar}}
\author[1]{\fnm{Evgenia} \sur{Klironomou}}
\author[1]{\fnm{Michela Florinda} \sur{Picardi}}
\author[1]{\fnm{Riccardo} \sur{Bertini}}
\author[1]{\fnm{Aleksandra} \sur{Deeva}}
\author*[1]{\fnm{Georgia T.} \sur{Papadakis} \email{georgia.papadakis@icfo.eu}}

\affil*[1]{\orgdiv{ICFO -- Institut de Ciencies Fotoniques}, \orgname{The
    Barcelona Institute of Science and Technology}, \orgaddress{\street{Avinguda
      Carl Friedrich Gauss, 3}, \city{Castelldefels}, \postcode{08860},
    \state{Catalonia}, \country{Spain}}}

\abstract{Chirality in the mid-infrared spectral range plays a crucial role
  across physical, chemical, and biological sciences, yet sources of chiral
  infrared light do not currently exist. Their development, using principles
  from the mature field of metamaterials, requires complex three-dimensional
  architectures that call for high-resolution lithography. We leverage the
  natural optical anisotropy found in several van der Waals crystals, for
  example \MoO, to demonstrate experimentally that its twisted bilayers break
  inversion-rotation symmetry and are thereby intrinsically chiral. Via direct
  thermal emission measurements of microscopic twisted bilayers, we demonstrate
  that these heterostructures generate chiral light through
  incandescence. Twisted configurations of van der Waals materials do not
  require any lithography, and offer a platform for large-scale chiral filters
  and thermal sources beyond conventional meta-architectures. }


\maketitle

\noindent Chirality is the geometric property that makes an object not
superimposable onto its mirror image through neither rotation nor
translation. It plays a crucial role in the development of life as we know
it~\cite{adamala2024s, devinsky2021s} and becomes relevant across various
disciplines in applied science and technology. For example, the fundamental
vibrational modes of various molecules, occurring primarily in the mid-infrared
(mid-IR) region of the electromagnetic spectrum, are highly sensitive to chiral
configurations~\cite{nafie2011}. This sensitivity is often leveraged in organic
chemistry and pharmaceuticals, where distinguishing between enantiomers --
molecules that are mirror images of each other -- is essential, since these can
exhibit distinct biological activity and therapeutic effects~\cite{reddy2004,
  smith2009ts}. Techniques like vibrational circular dichroism spectroscopy
utilize mid-IR light to differentiate enantiomers, offering vital insight into
their intrinsic configurations and purity~\cite{jahnigen2023acie, xu2023lsa}.

\begin{figure}
  \centering
  \includegraphics[width=12cm]{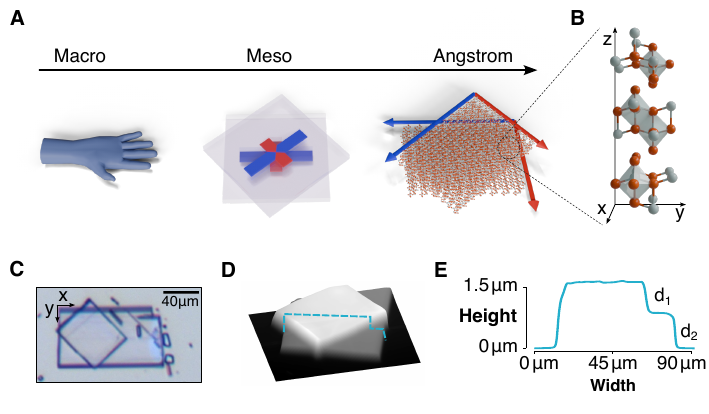}
  \caption{ 
    \textbf{Chirality and \MoO twisted bilayers.}
    \textbf{A}~Chiral objects across different scales. (Left) A human
    hand. (Middle) A simplified representation of twisted metamaterials and
    metasurfaces for artificial chirality. (Right) A twisted \MoO{} bilayer that
    enables intrinsic chirality. The \textbf{B}~Sketch of the atomic structure
    of \MoO{} showing the spatial coordinates \(xyz\) corresponding to the
    crystallographic directions $[100]$, $[001]$ and $[010]$,
    respectively~\cite{materials-project}.
    \textbf{C}~Microscope image of a twisted bilayer \MoO{} sample. The $x$- and
    $y$-axes of the crystal correspond to the bottom flake.
    \textbf{D}~AFM scan of the height profile of the sample shown \textbf{C}.
    \textbf{E}~Height profile along the dashed blue line in \textbf{D}, where
    thickness $d_{1}$ refers to the top flake and $d_{2}$ to the bottom flake.
  }
  \label{fig:fig1}
\end{figure}

Since sensing and detection of biological substances operate in the mid-IR
region, the role of IR light sources is critical in chiral analysis. However,
mid-IR lighting technology is primarily limited to the established -- but
lithographically complex -- quantum-cascade laser, or globars and Nernst glowers
that yield incoherent light. Both approaches lack the functionality to control
the polarization state of light in a compact device platform at mid-IR
frequencies. Conveniently, however, at near-room temperatures, the spectrum of
blackbody radiation emitted through incandescence peaks near 10\um, thereby
overlapping spectrally with the vibrational and phonon modes relevant in chiral
spectroscopy. Therefore, incandescence presents a promising avenue for
cost-effective mid-IR thermal sources~\cite{baranov2019nma}. Although light
generated from incandescence is by nature achiral, thermal emitters with
tailorable polarization characteristics have been reported in recent years,
leveraging the collective response of three-dimensional metamaterials and
metasurfaces~\cite{baranov2019nma}. In particular, most recently, thermal
emission with chiral characteristics has been reported with metamaterials
composed of elementary unit cells that are geometrically asymmetric in the
nanoscale~\cite{nguyen2023o, nolen2024nn}. However, the large-scale development
of such meta-architectures requires expensive high-resolution lithography and
synthesis, and their large-scale thermal excitation entails considerable
technical challenges as well.

\section{Chirality through twisting}

In Fig.~\ref{fig:fig1}A, we illustrate chirality across different scales,
starting with the most familiar chiral object in the macro-scale, a human hand,
which serves as an example of mirror asymmetry. Beyond the aforementioned
principles of conventional three-dimensional metamaterials with geometrically
asymmetric unit cells, even achiral objects, such as a geometric cross, can
serve as a unit cell of a planarized metamaterial that demonstrates chirality
through twisting adjacent unit cells~\cite{zhao2012nc, decker2009olo,
  zhao2017chirality}, thereby inducing various effects typically observed in
three-dimensional metamaterials with quasi-two-dimensional
motifs~\cite{gansel2009s, wang2009joapao, wang2016n}. As an example, the middle
of Fig.~\ref{fig:fig1}A shows a twisted combination of such crosses that break
inversion-rotation symmetry, representing the meso-scale and the regime of
metamaterials. The realization of planarized twisted metamaterials, however, is
subject to similar fabrication challenges as their three-dimensional
counterparts~\cite{nguyen2023o, nolen2024nn}.

In this work, we experimentally demonstrate a fundamentally different approach
to inducing chirality, leveraging the unique properties of emerging
low-dimensional van der Waals crystals. Instead of relying on external
structural modifications as in the case of metamaterials, we harness the
\textit{natural, intrinsic} in-plane optical anisotropy observed recently in
several emerging van der Waals materials, such as $\alpha$-molybdenum trioxide
(\MoO)~\cite{ma2018n, zheng2018am, alvarez-perez2020am} and $\alpha$-vanadium
pentoxide (\VO)~\cite{taboada-gutierrez2020nm}, and twist adjacent layers to
induce chirality. The concept of twisted anisotrpic bilayers is demonstrated on
the right of Fig.~\ref{fig:fig1}A, where twisted layers (heterostructures) of an
in-plane anisotropic material are shown. This can be understood in direct
analogy with the meso-scale and the case of twisted crosses; the blue and red
colors in the middle- and left-side of Fig.~\ref{fig:fig1}A, represent,
respectively, the two branches of a cross and the two dissimilar symmetry axes
of an in-plane anisotropic crystalline material. In both cases,
inversion-rotation symmetry is evidently broken via twisting. Importantly,
however, in the platform introduced here via twisting uniform anisotropic
crystals, chiral effects are enabled without the need for any patterned
meta-atoms or their coupling. The reported chirality is intrinsic and thereby
geometrically robust, as opposed to extrinsic chirality which is only observable
solely for certain orientations of incident light~\cite{plum2008apl}. The
reported chiral effects additionally remain present even in the case of twisted
monolayers, enabling two-dimensional chirality. The only requirement for
inducing chirality in this platform is a pronounced intrinsic in-plane
anisotropic material response. While several numerical studies have recently
explored twisting anisotropic materials for mid-IR and terahertz chirality
\cite{wu2022cpb, hou2024apl, lu2024ap, song2024o&lt, wang2024o&lt, wu2025oeo},
there has been no previous experimental demonstration.

\begin{figure}
  \centering
  \includegraphics[width=5.5cm]{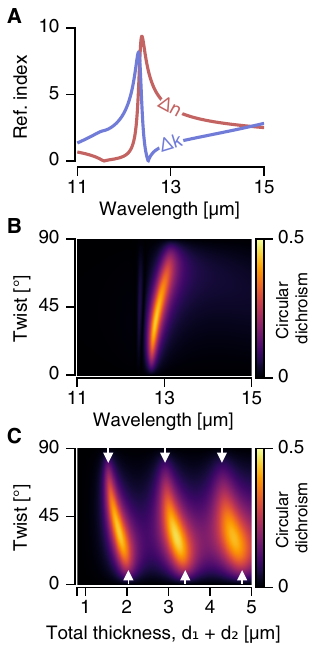}
  \caption{
    \textbf{Twisted \MoO{} bilayers for circular dichroism.}
    \textbf{A}~Linear birefringence $\Delta n$ and linear dichroism $\Delta k$ of a
    single \MoO{} flake.
    \textbf{B}~Transfer matrix simulation of a twisted bilayer composed of
    \MoO, with $d_1 = 0.8$\um{} and $d_{2} = 1$\um. Spectrum of circular dichroism
    as a function of the relative twist angle between the two adjacent \MoO{}
    layers.
    \textbf{C}~Dependence of circular dichroism on $d_2$, while $d_1 = 0.8\,\mu$m
    at a wavelength of 12.8\um. The $x$-axis shows the total thickness of the
    bilayer. White arrows indicate the total thickness for which the structure
    is resonant at twist angles of 0\degree{} (bottom ones) and 90\degree{} (top
    ones).
  }
  \label{fig:fig2} 
\end{figure}

We experimentally realize twisted bilayers composed of exfoliated flakes of
\MoO, which exhibit strong in-plane anisotropy at mid-IR
frequencies~\cite{alvarez-perez2020am}, and show via absorption measurements
that left- and right-hand-polarized light interacts differently with twisted
layers of \MoO, revealing a strong chiral response. For the first time, we carry
out direct thermal emission measurements of van der Waals-based twisted flakes
that have small lateral dimensions (tens of micrometers), and report thermally
excited chiral light in a lithography-free platform. These results pave the way
to simplified photonic functionalities leveraging the unique properties of
low-dimensional materials and eliminating the requirement of lithography for
chirality mid-IR engineering.

\section{Circular dichroism in twisted bilayers}
Owing to its interlayer interactions, \MoO{} is an ideal material for realizing
twisted heterostructures. Due to the orthorhombic structure of the \MoO{}
crystal~\cite{alvarez-perez2020am} (Fig.~\ref{fig:fig1}B), exfoliated ﬂakes
typically possess a rectangular shape, as shown in the microscope image of a
twisted bilayer in Fig.~\ref{fig:fig1}C. We define the crystal directions
$[100]$,$[001]$ and $[010]$, as the $x$-,$y$- and $z$-axis, respectively
(Fig.~\ref{fig:fig1}B). An atomic force microscopy (AFM) scan of the twisted
bilayer is presented in Fig.~\ref{fig:fig1}D, while the corresponding height
profile is shown in Fig.~\ref{fig:fig1}E. Henceforth, we denote the thicknesses
of the top and bottom flakes as $d_1$ and $d_2$, respectively.

We start with the case of a single flake of \MoO{}, the linear birefringence
$\Delta n$ and linear dichroism $\Delta k$ of which can be defined as the difference in
refractive indices and extinction coefficients along the two in-plane crystal
axes ($x$ and $y$). These properties are presented in Fig.~\ref{fig:fig2}A and
demonstrate the pronounced in-plane anisotropy of this material. Both $\Delta n$ and
$\Delta k$ exhibit a strong resonance near 12.3\um{} corresponding to the phonon mode
(Reststrahlen band) of \MoO{} occurring along the $x$-direction of its
crystal. Despite the strong anisotropy, a single flake of \MoO{} remains
intrinsically achiral. By contrast, twisting two adjacent layers of \MoO{}
introduces an intrinsic chiral response, as shown in Fig.~\ref{fig:fig2}B. This
chiral response can be described through circular dichroism (CD), which
quantifies the difference in the absorption between right- and left-circularly
polarized light. For a reflective substrate with vanishing transmission, CD can
be defined as:
\begin{equation}\label{eq:CD-ref}
  \text{CD} = \left|R_{\circlearrowright} - R_{\circlearrowleft}\right|,
\end{equation}
where $R_{\circlearrowright}$ and $R_{\circlearrowleft}$ denote the reflectance of right- and left-hand
circularly polarized light. The calculations of Fig.~\ref{fig:fig2}B are carried
out with the transfer matrix method~\cite{enders2024} at normal incidence, as a
function of the wavelength and the relative twist angle between adjacent \MoO{}
layers on a gold substrate. The layer thicknesses were set to $d_1 = 0.8$\um{}
and $d_2 = 1$\um, matching the dimensions of the devices used in the
experiments presented below. As shown in Fig.~\ref{fig:fig2}B, a maximum CD of
0.44 at a twist angle of approximately 36\degree{} is predicted in the spectral
range where \MoO{} exhibits maximal anisotropy (Fig.~\ref{fig:fig2}A). As
discussed earlier, this effect arises from the broken mirror symmetry introduced
by twisting the two layers relative to each other, and thus persists for all
twist angles between 0\degree{} and 90\degree. By contrast, at a twist angle of
0\degree, the crystals' axes of the two adjacent layers are aligned, making the
heterostructure a single uniform crystal of \MoO{} that preserves mirror
symmetry. Similarly, mirror symmetry is also preserved at a twist angle of
90\degree. Thereby, CD vanishes at the twist angles of 0\degree{} and 90\degree.

Although twisting adjacent layers suffices to break mirror symmetry for any
thicknesses of the layers, the chiral effect can be amplified by controlling the
interference between the layers. This is achieved by selecting the thicknesses
so that Fabry--Pérot resonances are supported (see Supplementary
Note~1). Fig.~\ref{fig:fig2}C illustrates this principle by 
varying the thickness of the bottom layer, $d_2$ while keeping the top layer's
thickness fixed at $d_1 = 0.8$\um. The $x$-axis represents the total bilayer
thickness, $d_1+d_2$. The bottom and top vertical white arrows indicate the
thicknesses for which absorption is maximized for $x$- and $y$-polarized light,
respectively, corresponding to twist angles of 0\degree and 90\degree, when the
$x$-axis of top layer of \MoO{} aligns with the $x$-axis and $y$-axis of the
bottom layer. In both cases, mirror symmetry is preserved as discussed above. By
contrast, CD becomes resonant at angles between 0\degree and 90\degree, for
various sets of thicknesses that can be analytically described (see
Supplementary Note~1). We note that, due to the resonant and very large optical
anisotropy of \MoO{} (Fig.~\ref{fig:fig2}A), the total thickness 
required for maximal CD remains sub-wavelength at mid-IR frequencies.

\section{Absorption spectroscopy}

\begin{figure}
  \centering
  \includegraphics[width=12cm]{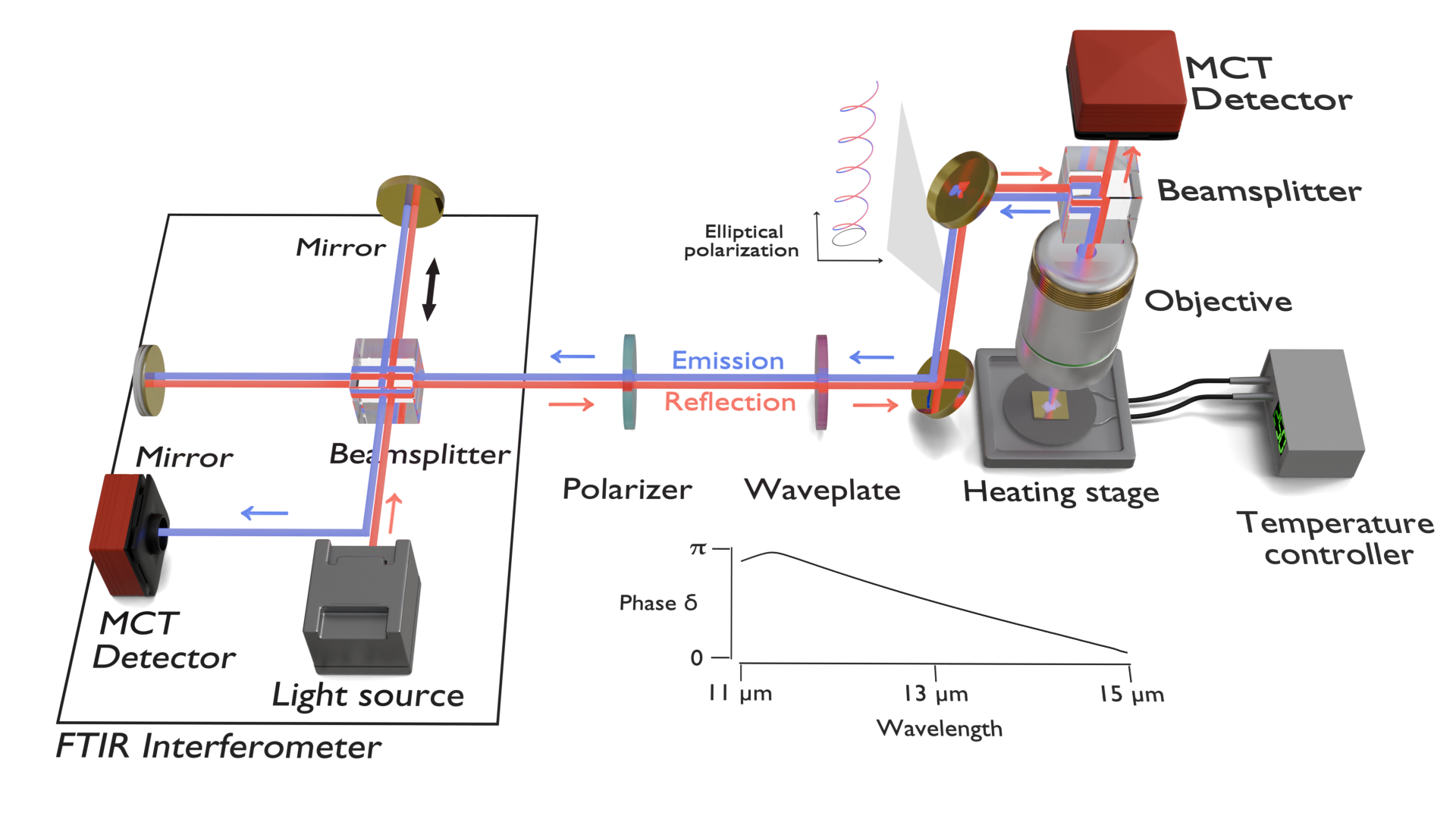}
  \caption{
    \textbf{Schematic of the experimental setup of the FTIR microscope for
      absorption and emission measurements.}
    The microscope operates in two modes: reflection and emission. In reflection
    mode (red beam path), infrared light from a source within the FTIR
    interferometer is directed onto the sample, and the reflected light is
    collected by a MCT detector inside the microscope. In emission mode (blue
    beam path), the sample's thermal emission is collected by the microscope
    objective, re-directed through the interferometer, and detected by another
    MCT detector. A polarizer and a waveplate are positioned between the
    microscope and the FTIR interferometer to analyze the polarization state of
    the light in both reflection and emission modes. The diagram below the
    wave-plate illustrates the phase shift ($\delta$) it introduces as a function of
    wavelength. 
  }
  \label{fig:fig3}
\end{figure}

\begin{figure}
  \centering
  \includegraphics[width=12cm]{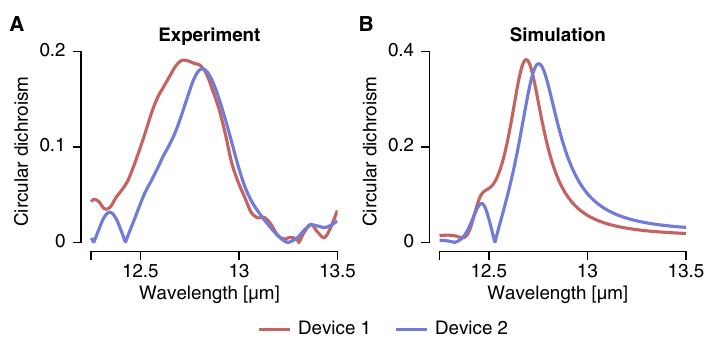}
  \caption{
    \textbf{Experimental measurements of CD in absorption.}
    \textbf{A}~FTIR microscope measurements of the CD spectrum of two twisted
    \MoO{} devices (\emph{Device~1} with $d_1 = 0.6$\um, $d_2 = 1.1$\um{} and twist angle 
    of 33\degree, \emph{Device~2} with $d_1 = 0.85$\um, $d_2 = 0.8$\um{} and twist angle of
    42\degree).
    \textbf{B}~Corresponding transfer matrix simulations for the two devices
    shown in \textbf{A}.
  }
  \label{fig:fig4} 
\end{figure}

We experimentally realized twisted bilayers of \MoO{} via exfoliation and
stacking (see Methods), aiming to operate near a maximum of CD based on the
predictions of Fig.~\ref{fig:fig2}C. The lateral dimensions of the fabricated
bilayers range from 10\um{} to 40\um. The two devices discussed henceforth are
termed \emph{Device~1} and \emph{2} and consist of a 0.6\um{}-thick flake on top
of a 1.1\um{}-thick flake twisted at an angle of 33\degree and a 0.8\um{}-thick
flake on top of a 0.85\um{}-thick flake, twisted at an angle of 42\degree{}
(Fig.~\ref{fig:fig1}C-E), respectively. Both devices were transferred onto a
gold-coated glass substrate to perform CD measurements in reflection. We
employed a Fourier transform infrared (FTIR) microscope ($36\times$ magnification,
Bruker, Hyperion~\rom{2}) for these measurements. The measured area of the
devices was restricted to that containing twisted flakes using the knife-edge
aperture of the FTIR microscope.

With respect to Eq.~\ref{eq:CD-ref}, determining the spectrum of CD requires a
broadband quarter-wave plate to circularly polarize the incident beam onto the
sample (red beam path in Fig.~\ref{fig:fig3}). Nonetheless, ideal broadband
quarter-wave plates are not commercially available at mid-IR
wavelengths~\cite{enders2024cm}. To circumvent this, we utilized a narrow-band
wave plate designed for operation near the wavelength of 13\um{} (CdSe from
VM-TIM GmbH), and characterized the phase shift that it introduces between the
$x$- and $y$-component of the electromagnetic field ($\delta(\lambda)$), for the whole
spectral range of interest (see diagram in Fig.~\ref{fig:fig3}) using the
technique outlined in~\cite{kilchoer2019ap}. The combination of this wave plate
and polarizer, which initially polarizes the IR light of the FTIR spectrometer
source (Bruker, Vertex~80), are shown in the beam path of
Fig.~\ref{fig:fig3}. Consistent with Eq.~\ref{eq:CD-ref}, it can be shown that,
for a reflective substrate with vanishing transmission, CD can also be expressed
as (see Supplementary Note~2):
\begin{equation}\label{eq:CD-conv}
  \text{CD} = \left|\frac{R_{45^{\circ}} - R_{-45^{\circ}}}{\sin(\delta)}\right|,
\end{equation}
where $R_{45^{\circ}}$ and $R_{-45^{\circ}}$ are the intensities of the reflected
electric field when the waveplate's fast axis is rotated by 45\degree{} and
$-45$\degree, respectively, with respect to the linearly polarized light exiting
the polarizer. Within the spectral range of interest, the phase shift of the
wave plate ($\delta$) presents values near 90\degree, ensuring that the factor
$\sin(\delta)$ remains near-unity, thereby avoiding singularities in
Eq.~\ref{eq:CD-conv} and minimizing measurement errors.

These FTIR measurements are shown in Fig.~\ref{fig:fig4}A, yielding nearly
20\,\% CD, confirming the predictions in Fig.~\ref{fig:fig2} as well as previous
theoretical estimations~\cite{wu2022cpb, hou2024apl, lu2024ap,song2024o&lt}. The
theoretical predictions for the experimentally measured devices are shown in
panel Fig.~\ref{fig:fig4}B, demonstrating the same spectral position where CD is
maximized, but larger values of CD as compared to the experimental
measurements are observed. This discrepancy arises from the small lateral dimensions of the
fabricated twisted areas of \MoO, which are on the scale of the measured
wavelength. In particular, transfer matrix calculations assume infinitely
extended surfaces, thereby they do not account for edge effects that arise in
small-area exfoliated samples. The slight difference between the resonant
position of CD in the two devices is expected since the thicknesses of the two
adjacent layers differ (see Fig.~\ref{fig:fig2}C).

The results of Fig.~\ref{fig:fig4} confirm that twisting of anisotropic bilayers
induces a strong chiral response. According to Kirchhoff's law of thermal
radiation~\cite{kirchhoff1860ap}, the preferential circular polarization
observed in absorption measurements (red beam path in Fig.~\ref{fig:fig3})
should also manifest in thermal emission measurements (blue beam path in
Fig.~\ref{fig:fig3}), which is confirmed in the following section.

\section{Emission spectroscopy}

\begin{figure}
  \centering
  \includegraphics[width=12cm]{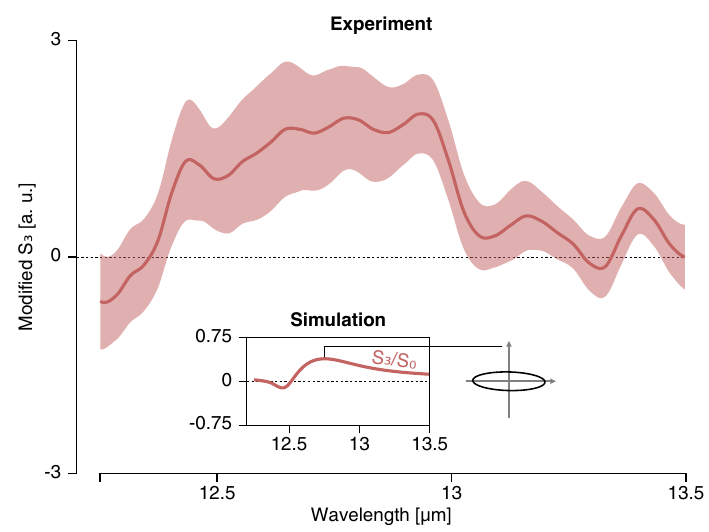}
  \caption{
    \textbf{Experimental measurements of CD in thermal emission.}
    Modified Stokes parameter $\widetilde{S}_3$ of \emph{Device~2} ($d_1 =
    0.85$\um, $d_2 = 0.8$\um{} and twist angle  of 42\degree) at a sample
    temperature of $300\,^{\circ}$C. The red shaded region demonstrates the error of
    the measurement arising from intensity fluctuations and angular 
    uncertainties inherent to the polarization state analyzer. The inset shows
    transfer matrix calculations of the normalized Stokes parameter
    $S_{3}/S_{0}$ and the polarization ellipse the wavelength of 12.7\um, where
    $S_{3}/S_{0}$ reaches its peak.
  }
  \label{fig:fig5} 
\end{figure}

We used the same experimental setup shown in Fig.~\ref{fig:fig3} for the thermal
emission measurements, directly measuring the radiation coming from the sample
itself. The sample was placed on a heating stage and its thermal radiation was
re-directed towards a mercury cadmium telluride (MCT) detector (see blue beam
path). Measuring the polarization state of thermal emission from microscopic
samples presents considerable experimental challenges. Most importantly, the
signal of the radiation emitted by a microscopic sample is several orders of
magnitude smaller than the background radiation from the surrounding and the
optical components in the beam path. This makes it difficult to isolate and
detect the sample's thermal signature. To ensure reliable measurements with
maximal signal, we focused on \emph{Device~2}, which had a significantly larger
twisted area compared to \emph{Device~1}, with lateral dimensions of
approximately 40\um{} (Fig.~\ref{fig:fig1}C). Additional challenges in the
measurement of thermal emission from microscopic samples include the detector
drift, which is on the same order of magnitude as the signal itself, as well as
heat-induced mechanical movement of the sample by tens of microns, introducing
instabilities. Furthermore, the sublimation temperature of molybdenum oxide is
$540\,^{\circ}$C~\cite{molina-mendoza2016cm}, thereby we operated at much lower
temperatures to prevent damage of the sample, however this further limited the
detectable signal. All aforementioned parameters affected measurement precision
and required accurate control of the instrumentation to obtain reliable data. A
detailed analysis of these factors is provided in Supplementary Notes~3 and 4.

The polarization state of the emitted radiation from the twisted bilayer \MoO{}
was characterized using Stokes polarimetry, employing a
polarization state analyzer -- a combination a linear polarizer and a waveplate --
following the method outlined by Nguyen \textit{et al.}~\cite{nguyen2023o} and
Sabatke \textit{et al.}~\cite{Sabatke:00}. The emitted radiation from the
twisted bilayers is partially polarized and can be fully described by the Stokes
vector $\mathbf{S} = (S_0, S_1, S_2, S_3)$, where $S_{0}$ represents the total
intensity of the emitted light, $S_{1}$ and $S_{2}$ quantify the difference in
intensity between horizontally and vertically polarized light and diagonally and
anti-diagonally polarized light, respectively, while $S_{3}$ characterizes the
difference in intensity between right- and left-circularly polarized light. The
quantity $S_{3}/S_{0}$ equals CD, making it the parameter of interest in
describing chiral properties of emitted radiation. Due to the small lateral
dimensions of the twisted area of the sample, the measured Stokes parameters
contained contributions from the entire microscope field of view that includes
the twisted area of the sample but also areas of the underlying single flake and
that of the bare substrate. Thereby, the component $S_{0}$ pertained to
unpolarized light since it accounted for the total emission from all areas of
the sample and the unpolarized background, while the linear polarization
components $S_{1}$ and $S_{2}$ had contributions from both the twisted region
and the underlying single flake. By contrast, since neither the underlying
single \MoO{} flake nor the substrate exhibit chiral emission, $S_{3}$, the
circular polarization component, arised only from the twisted region which is
the only one that breaks inversion-rotation symmetry thereby emitting chiral
light.

FTIR spectroscopy records relative intensity, normalizing the sample's spectrum
of thermally emitted light to that of a reference sample. We selected the gold
substrate as the reference surface, since it is both achiral and spectrally flat
across the mid-IR range. As a result, the measured Stokes vector was normalized
to the total unpolarized reference signal that includes the background signal
from the instrument and beam path, $B_{\text{Reference}}$, which is also
spectrally flat within the spectral range of interest. Thereby, the measured
Stokes vector can be written as:
$\widetilde{\mathbf{S}} = {\mathbf{S}}/({I_{\text{Reference}} +
  B_{\text{Reference}}})$. The influence of $B_{\text{Reference}}$ can be
mitigated by carrying out measurements at different
temperatures~\cite{xiao2020l&pr}, however this was not feasible in our setup due
to detector drift, requiring short timescales in the experimental
measurements. The thermal emission measurements that we performed were taken at
$300\,^{\circ}$C.

Fig.~\ref{fig:fig5} presents experimental measurements of the normalized Stokes
parameter $\widetilde{S}_{3}$ characterizing circularly polarized emission:
\begin{equation}
  \widetilde{S}_{3} = \frac{S_{3}}{I_{\text{Reference}} + B_{\text{Reference}}}.
\end{equation}
The quantity $\widetilde{S}_{3}$ serves as a practical and well-defined
indicator of chiral thermal emission in our system under the imposed
experimental constraints. In particular, due to the absence of spectral features
in $I_{\text{Reference}}$ and $B_{\text{Reference}}$, the spectral shape and
strength of $\widetilde{S}_{3}$ reliably depicts the spectral features of
$S_{3}$. As expected from Kirchhoff's law, the spectrum in Fig.~\ref{fig:fig5}
is also in quantitative agreement with the chiral features in the absorption
measurements of Fig.~\ref{fig:fig4}. The inset in Fig.~\ref{fig:fig5}shows
transfer matrix calculations of the normalized Stokes parameter $S_{3}/S_{0}$
for a twisted \MoO{} bilayer, also showing quantitative agreement with the
measurement. By considering the calculated linear components of the Stoke's
vector, $S_{1}$ and $S_{2}$ (see Supplementary Fig.~S8), we also draw in the
inset of Fig.~\ref{fig:fig5} the polarization ellipse at the wavelength where
$S_{3}$ reaches its peak. As expected, in contrast to conventional blackbody
thermal emission that is unpolarized, the emitted radiation from the twisted
bilayer exhibits a pronounced elliptical polarization component. The red shaded
region in Fig.~\ref{fig:fig5} represents the error of the measurement,
accounting for intensity fluctuations and angular uncertainties inherent to the
polarization state analyzer.

\section{Conclusion and outlook}

In this work, we showed experimentally that homogeneous flakes of van der Waals
materials with an in-plane anisotropy can serve as a platform for engineering
intrinsic chirality and generating chiral light. Without loss of generality, we
demonstrated these properties with \MoO~\cite{ma2018n, zheng2018am,
  alvarez-perez2020am}, however the effect is general and can be observed with
other low-dimensional materials as well, for example
\VO~\cite{taboada-gutierrez2020nm}. Both \MoO{} and \VO{} as well as their
heterostructures have been recently reported to demonstrate various optical
phemomena that originate predominantly from the directional nature of surface
phonon polaritons that occur in both materials at mid-IR
frequencies~\cite{chen2020nm, dai2020ncb, hu2020n, duan2021sa,
  alvarez-perez2022sa, guo2023nc, hu2023s, sternbach2023s}. By contrast, the
intrinsic mid-IR chirality and thermally generated chiral light reported here
\textit{do not rely} on polaritonic waves and can be observed in \textit{any}
twisted configuration of in-plane anisotropic materials that have sufficient
optical losses.

Although thermal emission, in other words incandescence, is by nature
unpolarized, unidirectional and incoherent, we demonstrated that twisted,
unpatterned bilayers of \MoO{} can dramatically modify the characteristics of
blackbody radiation, emitting chiral thermal radiation. We achieved this by
fabricating twisted \MoO{} bilayers with exfoliation and stacking. Despite
challenges related to conducting thermal emissivity measurements of samples with
microscopic lateral dimensions and at low-temperatures, we were able to detect
and confirm the circular polarization state of the emitted light. This
highlights the robustness of twisted bilayer devices and their potential for
applications in mid-IR polarization control, mid-IR lighting, chiral sensing and
detection. The realization of twisted bilayers does not require \textit{any}
lithography, thereby introducing a simple, planar, and scalable roadmap for
chirality engineering beyond the regime of traditional metamaterials and
lithography-based meta-devices. The reported values of circular dichroism are
primarily limited by the finite lateral dimensions of the flakes rather than
intrinsic material constraints; therefore, larger-area fabrication approaches
such as chemical vapor deposition can help to achieve even stronger chiral
signatures.

\bmhead{Methods}

We mechanically exfoliated flakes of \MoO{} with polydimethylsiloxane-based
exfoliation and transfer (X0 retention, DGL type from Gelpak) at
$90\,^{\circ}$C~\cite{castellanos-gomez20142m}. Firstly, the bottom flake was
transferred onto gold-coated (150\,nm) glass and consequently the top flake was
transferred onto the bottom flake at the desired twist angle, using an optical
microscope which enables rotation and positioning of the flakes. The dielectric
permittivity of the batch of \MoO{} (2D Semiconductors, Bridgman growth
technique) used in the devices presented in this article was extracted using
FTIR spectroscopy, following the method described in~\cite{sarkar2023}. These
permittivity values serve as the basis for our transfer matrix simulations.

We conducted FTIR micro-spectroscopy measurements using a Bruker
Hyperion~\rom{2} microscope coupled with a Bruker Vertex~80 FTIR spectrometer 
equipped with a MCT detector. A \(\times36\) Cassegrain objective was
employed for collection. We utilized two linear ZnSe holographic wire
grid polarizers from Thorlabs and CdSe waveplates from VM-TIM GmbH. The size
of the measured area was controlled using the knife-edge aperture of the
microscope.

\bmhead{Acknowledgements} We acknowledge fruitful discussion with Dr.\ Krystian
Nowakowski, Dr.\ Hanan Herzig Sheinfux and Prof.\ Frank Koppens in the Quantum
Nano-Optoelectronics Group at ICFO and the generous sharing of several optical
components. We also acknowledge fruitful discussions with Prof.\ T. Peter
Rakitzis (University of Crete, Foundation for Research and Technology Hellas),
Prof.\ Lisa V. Poulikakos (University of California San Diego) and Dr.\ Ivan
Fernandez Gorbaton (Karlsruhe Institute of Technology). We acknowledge Ryo
Mizuta Graphics for the provision of optical component assets utilized in
Figure~3.

\section*{Declarations}
\bmhead{Funding}
MTE acknowledges support from MCIN/AEI/10.13039/501100011033 \newline (PRE2020-094401)
and FSE ``El FSE invierte en tu futuro''. R.B. acknowledges funding from the
European Union’s Horizon 2020 research and innovation programme under the Marie
Skłodowska-Curie grant agreement no. 847517. MFP Acknowledges support from the
Optica Foundation 20th Anniversary Challenge Award. MFP and GTP received the
support of fellowships from ``la Caixa'' Foundation (ID 100010434). The
fellowship codes are LCF/BQ/PI23/11970026 and LCF/BQ/PI21/11830019. GTP also
acknowledges support from the Spanish MICINN (PID2021-125441OA-I00,
PID2020-112625GB-I00, and CEX2019-000910-S), Generalitat de Catalunya (2021 SGR
01443), Fundació Cellex, and Fundació Mir-Puig.
\bmhead{Competing interests}
There are no competing interests to declare.
\bmhead{Data and materials availability}
All data are available in the manuscript or the supplementary materials.
\bmhead{Code availability}
The code used for the TMM simulations is available at
\url{https://github.com/mtenders/GeneralizedTransferMatrixMethod.jl}
\bmhead{Author contributions}
Following the CRediT taxonomy:\newline
Conceptualization: GTP. \newline
Funding acquisition: GTP. \newline
Investigation: MTE, EK, MS, RB, AD. \newline
Formal Analysis: MTE, EK, MS, MFP. \newline
Methodology: MTE, EK. \newline
Resources: GTP. \newline
Software: MTE. \newline
Visualization: MTE. \newline
Writing -- original draft: MTE. \newline
Writing -- review \& editing: MTE, EK, MS, MFP, RB, AD, GTP.

\bibliography{references}

\end{document}